\documentclass[12pt]{iopart}
\makeatletter
\def\@dotsep{4.5}
\makeatother

\usepackage[ps2pdf]{hyperref}
\hypersetup{
  bookmarksnumbered=true,
  colorlinks,
  pdfauthor={Daniele Coslovich},
  pdfkeywords={molecular dynamics glass transition network liquids energy landscape},
  citecolor={blue},
  linkcolor={blue},
  urlcolor={magenta},
}

\usepackage{graphicx,iopams}

\newlength{\onefig}
\newlength{\twofig}
\newcommand{\lav}{\langle\,}
\newcommand{\rav}{\,\rangle}
\newcommand{\bq}{\begin{eqnarray}}
\newcommand{\eq}{\end{eqnarray}}
\newcommand{\bqn}{\begin{eqnarray*}}
\newcommand{\eqn}{\end{eqnarray*}}
\newcommand{\beq}{\begin{equation}}
\newcommand{\eeq}{\end{equation}}
     
\newcommand{\der}[2]{\frac{d #1}{d #2}}    
\newcommand{\eps}{\epsilon}
\newcommand{\dt}{\delta t}

\newcommand{\ve}[1]{\boldsymbol{#1}}

\begin{document}
  
\title[Dynamics and energy landscape in a tetrahedral network
  glass-former]{Dynamics and energy landscape in a tetrahedral network
  glass-former: Direct comparison with models of fragile liquids}

\author{D. Coslovich}
\address{Insitut f\"ur Theoretische Physik and CMS, Technische Universit\"at
  Wien -- Wiedner Hauptstra{\ss}e 8-10, A-1040 Wien, Austria} 
\ead{coslovich@cmt.tuwien.ac.at}

\author{G. Pastore}
\address{Dipartimento di Fisica Teorica, Universit{\`a} di Trieste
  -- Strada Costiera 11, 34014 Trieste, Italy} 
\address{CNR-INFM Democritos National Simulation Center -- 
  Via Beirut 2-4, 34014 Trieste, Italy} 
\ead{pastore@ts.infn.it}

\date{\today}

\begin{abstract}
We report Molecular Dynamics simulations for a new model of
tetrahedral network glass-former, based on short-range, spherical
potentials. Despite the simplicity of the forcefield employed, our
model reproduces some essential physical properties of silica, an
archetypal network-forming material. Structural and dynamical
properties, including dynamic heterogeneities and the nature of local
rearrangements, are investigated in detail and a direct comparison
with models of close-packed, fragile glass-formers is performed. The
outcome of this comparison is rationalized in terms of the properties
of the Potential Energy Surface, focusing on the unstable modes of the
stationary points. Our results indicate that the weak degree of
dynamic heterogeneity observed in network glass-formers may be
attributed to an excess of localized unstable modes, associated to
elementary dynamical events such as bond breaking and reformation. On
the contrary, the more fragile Lennard-Jones mixtures are
characterized by a larger fraction of extended unstable modes, which
lead to a more cooperative and heterogeneous dynamics.
\end{abstract}
\pacs{61.43.Fs, 61.20.Lc, 64.70.Pf, 61.20.Ja}

\maketitle

\section{Introduction}
Network-forming amorphous materials are of great interest for
technological applications, as well as of fundamental importance for
the theoretical understanding of the glass transition. At a
microscopic scale, the structure of network glass-formers, in both the
amorphous and highly viscous regime, is characterized by strong
chemical ordering and atomic arrangements that usually form an open
tetrahedral network. Upon cooling from high temperature, transport
coefficients and structural relaxation times $\tau$ of network liquids
display a mild temperature dependence, often describable by the
Arrhenius law $\tau\approx\tau_{\infty}\exp(E/T)$. Network glass-formers
are thus ``strong'' in the Angell's classification
scheme~\cite{angell88}. In contrast, other classes of glass-formers,
including molecular, polymeric, and bulk metallic liquids, show
super-Arrhenius temperature dependence of $\tau$, i.e., ``fragile''
behaviour.

Ever since the introduction of the Angell's classification, the nature
of the distinction between fragile and strong liquids has been highly
debated. While the degree of fragility of a liquid correlates
\textit{quantitatively} with other macroscopic physical properties,
the existence of \textit{qualitative} differences between strong and
fragile systems has been questioned. Evidence of a
``fragile-to-strong'' crossover in simulated network
liquid~\cite{saika-voivod_fragile-to-strong_2001} and numerical
investigations of dynamic heterogeneities in model
glass-formers~\cite{vogel04a} suggest that network liquids may just be
an extreme case of the class of fragile systems~\cite{berthier07a}. In
contrast, theoretical work on kinetically constrained models of glassy
dynamics~\cite{garrahan03a} indicates that strong behaviour may arise
from the different nature of dynamical constraints. Moreover, the
energy landscape description of supercooled
liquids~\cite{stillinger95,debenedetti01a} shows that the organization
and connectivity of stationary points in the Potential Energy Surface
(PES) may be qualitatively different in fragile and strong
glass-formers. In this two latter scenarios, fragile and strong
liquids would thus belong to different ``universality classes'' of
glass-formers.

Silica is often considered as a prototypical network glass-former. In
recent years, several authors have studied structural and dynamical
properties of this system through numerical simulations, employing
both Molecular Dynamics (MD) and Monte Carlo techniques.  One of the
most realistic and popular models of silica available for molecular
simulations is the BKS model introduced by Van
Beest~\etal~\cite{vanbeest90}. In this model, the interaction between
Si and O atoms is described by a long-ranged Coulombic interaction,
plus a short range repulsion of the Born-Mayer type. Various physical
aspects of the supercooled and glassy regime of the BKS model have
been analysed, including the phase
diagram~\cite{saikavoivod00a,saikavoivod04a},
structural~\cite{vollmayr96a},
dynamical~\cite{horbach99a,horbach01a,vogel04a,berthier07a}, and
vibrational~\cite{taraskin_nature_1997,taraskin_anharmonicity_1999,benoit02a}
properties. Investigations of the energy landscape of the BKS model
have also been
performed~\cite{jund_computer_1999,la_nave_configuration_2002,bembenek_instantaneous_2001,saksaengwijit04a}.
Because of the long-ranged nature of the interactions, however,
simulations using the BKS model are computationally demanding. Hence,
development of simpler force-fields, capturing the basic features of
network liquids, is highly desirable. Recently, in fact, simplified
models for silica have been proposed, including short-ranged variants
of the original BKS potential~\cite{kerrache05a,carre07a} and
primitive models based on patchy
interactions~\cite{ford04a,demichele06a}. Other models of tetrahedral
network liquids (not directly related to silica) based on spherical,
patchy interactions have also been studied
recently~\cite{zaccarelli_spherical_2007} and in the
past~\cite{ferrante89a}.

In this work, we present a new model of network glass-former, based on
spherical, short-ranged potentials. Our model allows efficient
simulations and can be tuned to reproduce some relevant properties of
amorphous silica. It does not aim at a realistic description of liquid
and amorphous silica, yet it captures to a good extent the essential
physics of network glass-formers. Moreover, being able to describe
both network and ``simple'' glass-forming
liquids~\cite{coslovich07a,coslovich07b} with similar efficiency via
the same family of interactions, we can get an unusually detailed and
systematic comparison between the microscopic origins of their
structural relaxation.
In particular, we trace back the distinct dynamic features of network
glass-formers (e.g. strong behaviour, weak dynamic heterogeneity, bond
breaking and reformation processes) to the properties of the PES,
contrasting our findings with the case of the more fragile,
close-packed Lennard-Jones (LJ)
mixtures~\cite{coslovich07a,coslovich07b}. Our results emphasize the
role of the unstable modes of the PES, as a key to rationalize the
different dynamic behaviours of glass-forming liquids.

The paper is organized as follows: in section~\ref{sec:model} we
introduce our model of network glass-former; in  
section~\ref{sec:results} we describe its structural and dynamical
properties, while in section~\ref{sec:pes} we analyse the properties
of the stationary points of the PES, focusing on the unstable
modes. Finally, in section~\ref{sec:conclusions} we draw our
conclusions.

\section{Model}\label{sec:model}

Our model of network glass-former, called NTW herein, is a binary mixture of
classical particles interacting through the following forcefield
\begin{eqnarray}\label{eqn:u}
u_{\alpha\alpha}(r)&=&4\eps_{\alpha\alpha}\left(\frac{\sigma_{\alpha\alpha}}{r}\right)^{12} \\
u_{\alpha\beta}(r)&=&4\epsilon_{\alpha\beta}\left[ {\left(
    \frac{\sigma_{\alpha\beta}}{r} \right)}^{12} -
  {\left( \frac{\sigma_{\alpha\beta}}{r} \right)}^6 \right] \quad \alpha\neq\beta
\end{eqnarray}
where $\alpha$,$\beta$=1,2 are indexes of species. In the following,
we will use $\sigma_{11}$, $\epsilon_{11}$ and
$\sqrt{m_1\sigma_{11}^2/\epsilon_{11}}$ as reduced units of distance,
energy and time respectively. Keeping an eye on silica, we identify
large particles (species 1) with Si atoms and small particles (species
2) with O atoms, and we fix the number concentrations at $x_1=0.33$,
$x_2=0.67$. We also use the same mass ratio of O and Si atoms:
$m_2/m_1=0.57$. A smooth cut-off scheme is used to ensure continuity
of $u_{\alpha\beta}(r)$ at $r=2.2\sigma_{\alpha\beta}$ up to the
second derivative~\cite{grigera02}. The size of the samples considered
in this work is $N=N_1+N_2=500$. The presence of finite size effects
have been checked through simulations of larger systems ($N=2048$,
$8000$) and will be briefly discussed in section~\ref{sec:results}. We
performed MD simulations in the NVE ensemble using quenching protocols
and equilibration criteria similar to the ones of previous simulations
of LJ mixtures~\cite{coslovich07a,coslovich07b}. Equilibration and
production runs were performed using Berendsen rescaling and
velocity-Verlet algorithm, respectively. The time step $\dt$ was
varied between 0.001 (at high $T$) and 0.004 (at low $T$). The absence
of major systematic aging effects was checked by comparing
thermodynamic, structural, and dynamical properties in different parts
of the production runs. At the lowest temperatures, simulations
involved up to $3.5 \times 10^7$ and $7 \times 10^7$ steps for the
equilibration and production runs, respectively. Thanks to the short
range of the potentials and to the open local structure of the system,
these long runs took a few days on a 3.4 GHz Xeon processor.

\begin{figure}[tb]
\begin{center}
\includegraphics*[width=\twofig]{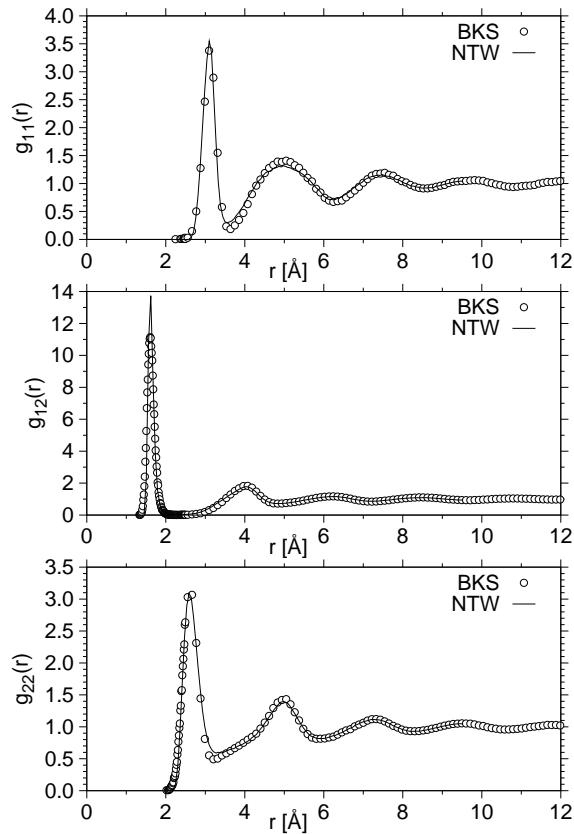}
\end{center}
\caption{\label{fig:gr} Partial pair correlation functions
  $g_{\alpha\beta}(r)$ for the NTW model (solid lines) and the BKS
  model for silica (open points). The thermodynamic state shown for
  BKS silica is $\rho=2.37 \textrm{g}/\textrm{\AA}^3,T=2750\textrm{K}$
  and the one of NTW is $\rho=1.655,T=0.39$ in reduced units. The BKS
  data refer to the MD simulations by Horbach and
  Kob~\cite{horbach96a}.  }
\end{figure}

To reproduce the open, tetrahedral local structure of network glasses, two
main physical ingredients must enter in the forcefield of our model: highly 
non-additive interaction radii and strong attraction between unlike
species. Building on previous experience~\cite{ferrante89a}, we determined the
following optimal set of interaction parameters:
\begin{eqnarray*}
 \sigma_{12}/\sigma_{11} = 0.49 &\quad &\sigma_{22}/\sigma_{11}=0.85 \\
 \eps_{12}/\eps_{11}     = 6.00 &\quad & \eps_{22}/\eps_{11}   =1.00
\end{eqnarray*}
To optimize the parameters above, we performed a series of preliminary
simulations at reduced density $\rho=\rho_{expt}\approx 1.53$, which
corresponds to the density of amorphous silica in normal experimental
conditions. The parameters were adjusted by requiring that the ratio
between the positions of the first peaks in $g_{12}(r)$ and
$g_{11}(r)$ was equal to that of Si-O and Si-Si interatomic distances
of amorphous silica. We also checked that at low temperature the
average coordination numbers, obtained from the integral of the radial
distribution functions, were close to the ideal tetrahedral ones,
i.e., $Z_{12}=4$, $Z_{21}=2$.

\begin{figure}[tb]
\begin{center}
\includegraphics*[width=\twofig]{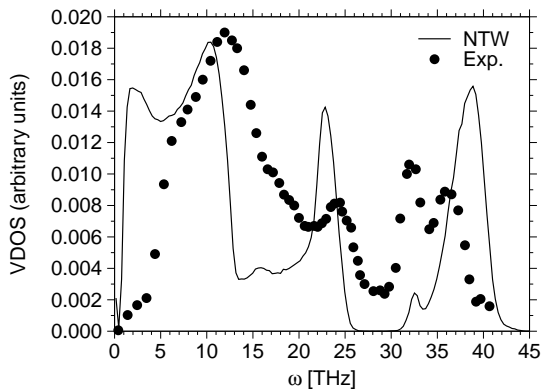} 
\end{center}
\caption{\label{fig:vdos} Vibrational density of states (VDOS)
  obtained from local minima of the potential energy at $\rho=1.53$,
  $T=0.30$ for the NTW model (solid line), compared to experimental
  VDOS of amorphous silica (filled points,
  from~\cite{carpenter85a}). The experimental data are convoluted with
  a correction function that alters their features only
  quantitatively~\cite{benoit02a}.}
\end{figure}

The physical units of our model can be be fixed to reproduce some
relevant properties of experimental and realistic models of silica,
such as the BKS model. We fixed the length scale $\sigma_{11}$ so that
the position of the first peak of $g_{11}(r)$ matched the mean Si-Si
distance (i.e., $3.12 \textrm{\AA}$) of amorphous silica. In this way
we obtained $\sigma_{11}=2.84\textrm{\AA}$.  To fix the energy scale
$\eps_{11}$, we compared the shape of the radial distribution
functions $g_{\alpha\beta}(r)$ to the ones obtained by Horbach and
Kob~\cite{horbach96a} for BKS silica at the state point
$\rho=2.37\textrm{g}/\textrm{\AA}^3$, $T=2750 \textrm{K}$ (see
figure~\ref{fig:gr}). The corresponding density in reduced units is
$\rho=1.655$. A good overall agreement of the liquid structure is
found around $T=0.39$, from which we estimate $\eps_{11}\approx
7000\textrm{K}$. Finally, we fixed the time scale of our model by
adjusting the mass scale $m_1$ so as to reproduce typical vibrational
frequencies of amorphous silica. Following previous studies (see for
instance~\cite{taraskin_nature_1997,taraskin_anharmonicity_1999,benoit02a}),
we determined the vibrational density of states (VDOS) through
diagonalization of the dynamical matrix calculated at local minima of
the potential energy at $\rho=1.53$, $T=0.30$. The choice
$m_1=8.7\times 10^{-23}\textrm{g} \approx 1.9 m_{Si}$ yields
reasonable agreement between the VDOS of our model and the
experimental VDOS of amorphous silica~\cite{carpenter85a} (see
figure~\ref{fig:vdos}, which is further discussed in
section~\ref{sec:results}). From the value of $m_1$ given above we
obtain the time unit
$t_0=\sqrt{m_1\sigma_{11}^2/\epsilon_{11}}=2.0\times
10^{-13}\textrm{s}$.

\section{Structure and dynamics}\label{sec:results}
In this section we further validate our model by analysing its structural and
dynamical properties. Our simulations spanned a wide range of densities:
$1.250\le\rho\le2.300$. At higher density ($\rho=2.800$) we found clear signs
of crystallization of our samples, but we did not attempt to determine
the crystallographic structure. At lower densities ($\rho \leq 1.250$) large
voids are formed in the network structure, and liquid-gas phase
separation might occur. In the following, we will mostly focus on the isochore
$\rho=1.655$, which corresponds to the density employed in several simulations
of BKS silica, at temperatures in the range $0.29\leq T \leq 1.50$.

\subsection{Structure and vibrations}\label{sec:structure}

The fact that the radial distribution functions of our model agree
rather well with those of the more realistic BKS model (see
figure~\ref{fig:gr}) and the overall qualitative shape of the VDOS
(see figure~\ref{fig:vdos}), already suggest that the NTW model should
capture some relevant physical aspects of network glass-formers, at
least for densities and temperatures where tetrahedral local ordering
is more pronounced. In this section we study in more details the
structural and vibrational properties of our model.

\begin{figure}[tb]
\begin{center}
\includegraphics*[width=\twofig]{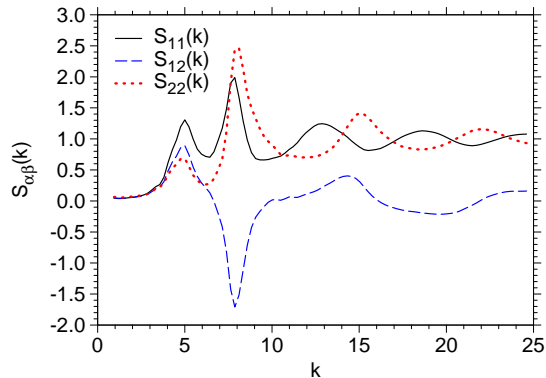} 
\end{center}
\caption{\label{fig:sk} Partial structure factors $S_{11}(k)$ (solid
  line), $S_{12}(k)$ (dashed line), and $S_{22}(k)$ (dotted line) at
  $\rho=1.655$ and $T=0.29$.}
\end{figure}

In figure~\ref{fig:sk} we show the partial structure factors
$S_{\alpha\beta}(k)$ obtained at the lowest temperature attained in
equilibrium conditions for $\rho=1.655$. The pre-peak (also called
first sharp diffraction peak) at $k\approx 5.0$ in $S_{11}(k)$ and
$S_{22}(k)$ signals the formation of intermediate range order. This is
a typical feature apparent at low temperature in network liquids. The
positions of the pre-peak and main peak ($k\approx 8.0$) in
$S_{11}(k)$ are in good agreement with those of $S_{SiSi}(k)$ in
amorphous silica and simulated BKS silica~\cite{horbach99a}.

\begin{figure}[tb]
\begin{center}
\includegraphics*[width=\twofig]{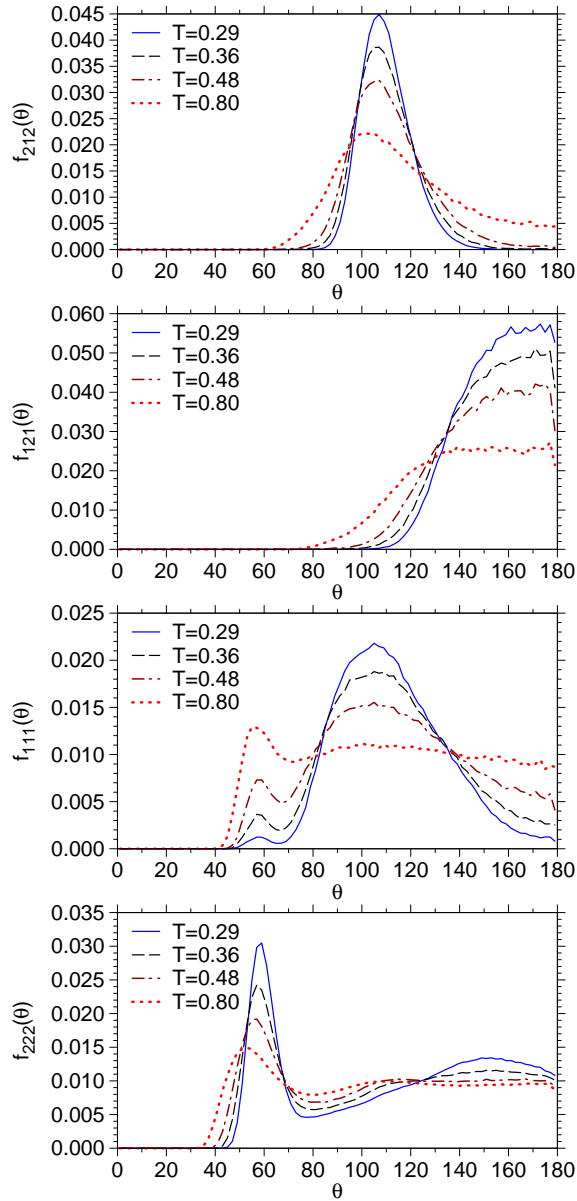} 
\end{center}
\caption{\label{fig:angle} Angular distribution functions
  $f_{\alpha\beta\gamma}(\theta)$ for $T=0.80$ (dotted lines), 0.48
  (dash-dotted lines), 0.36 (dashed lines), 0.29 (solid lines) at
  density $\rho=1.655$.}
\end{figure}

Further insight into the structural properties of the NTW model is
provided by the distribution $f_{\alpha\beta\gamma}(\theta)$ of angles
formed by a central particle of species $\beta$ with neighbours of
species $\alpha$ and $\gamma$, where
$\alpha,\beta,\gamma=1,2$. Particles of species $\alpha$ and $\gamma$
are considered neighbours if their distance is less than the minimum of
the radial distribution function at the corresponding $T$. The
normalized angular distribution functions $f_{121}(\theta)$ and
$f_{212}(\theta)$, shown in figure~\ref{fig:angle} for a few selected
temperatures, reveal the typical features associated to local
tetrahedral ordering. The broad peak in $f_{212}(\theta)$, located
around $\theta=108^{\circ}$, reflects the presence of slightly
distorted tetrahedra centered around particles of species 1. The
$f_{121}(\theta)$ shows a peak around $180^\circ$, which corresponds
the links formed by particles of species 2 connecting adjacent
tetrahedra. Note that the peak positions and the overall shape of
these distribution functions change only mildly below $T\approx
0.50$. Thus, below this temperature, which we will identify in the
next section as the onset temperature of
slow-dynamics~\cite{sastry98}, the NTW model displays a strong degree
of tetrahedral local ordering.

The angular distribution functions $f_{111}(\theta)$ and $f_{222}(\theta)$
(see two lower plots in figure~\ref{fig:angle}) provide information about the
intermediate range order of our model. The $f_{111}(\theta)$ displays a broad
peak located at $\theta\approx 105^{\circ}$, associated to distorted
corner-sharing tetrahedra. The smaller peak around $60^\circ$, due to
three-fold rings~\cite{jin94a}, decreases in height upon lowering the
temperature. At higher density ($\rho=2.300$, not shown here) this small peak
increases in intensity (at fixed $T$), while the peak at $\theta\approx
105^{\circ}$ splits in two sub-peaks. Similar variations upon compression were
found in the angular distribution functions of a more realistic model of
silica~\cite{jin94a}. Hence, we conclude that our simple model is able to
capture some non-trivial structural features of network glass-formers.

\begin{figure}[tb]
\begin{center}
\includegraphics*[width=\twofig]{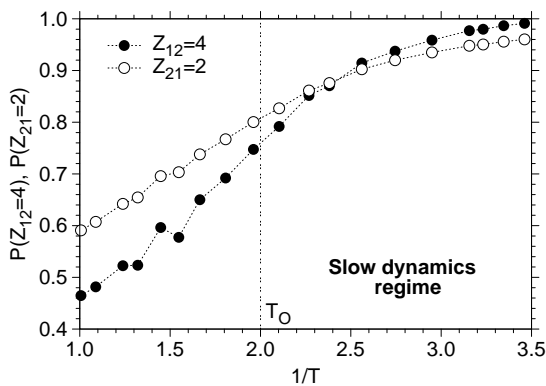} 
\end{center}
\caption{\label{fig:bonds} Fraction of particles with ideal
  tetrahedral coordination as a function of $1/T$: $P(Z_{12}=4)$
  (empty circles) and $P(Z_{21}=2)$ (filled circles) are shown for
  particles of species 1 and 2, respectively. The vertical dotted line
  marks the onset of the slow-dynamics regime.}
\end{figure}

To highlight the formation of a nearly ideal tetrahedral network at
low $T$, we show in figure~\ref{fig:bonds} the $T$-dependence of the
fraction of particles with ideal coordination numbers, i.e.,
$P(Z_{12}=4)$ and $P(Z_{21}=2)$ for particles of species 1 and 2,
respectively. To identify neighbouring particles we used the same
criterion as for the angular distribution functions discussed
above. The fraction of ideally coordinated particles is already substantial
around $T\approx 0.5$ [$P(Z_{12}=4), P(Z_{21}=2) > 0.70$] and
approaches unity at low $T$. This provides further indication that for
the density considered here the system is indeed in the optimal region
of network formation~\cite{zaccarelli_spherical_2007}.

A closer inspection of figure~\ref{fig:vdos} shows that the VDOS of
the NTW model reproduces all the qualitative features of the
experimental VDOS of amorphous silica. The relative positions of the
peaks in the VDOS of NTW match well enough those of the experimental
VDOS. Note that the absence of a peak at small frequencies ($\omega
\approx 4$ THz) in the experimental data is due to insufficient
experimental resolution~\cite{benoit02a,wischnewski98a}. A careful
comparison of simulated and experimental VDOS of silica can be found
in~\cite{benoit02a}. Here we only recall that the VDOS of the BKS
model is somewhat unrealistic at low and intermediate
frequencies~\cite{benoit02a}. Similar deficiencies have also been
found in recent modifications of the original BKS model employing
short-ranged potentials~\cite{carre07a}. Specifically, the distinct
peaks at 12 and 24 THz, as well as the small peak around 18 THz ($D_2$
line), are missing in the VDOS of BKS silica. Given the simplicity of
the forcefield employed, the success of the NTW model in reproducing
the main qualitative vibrational features of amorphous silica is
rather remarkable.

\subsection{Relaxation dynamics}

\begin{figure}[tb]
\begin{center}
\includegraphics*[width=\twofig]{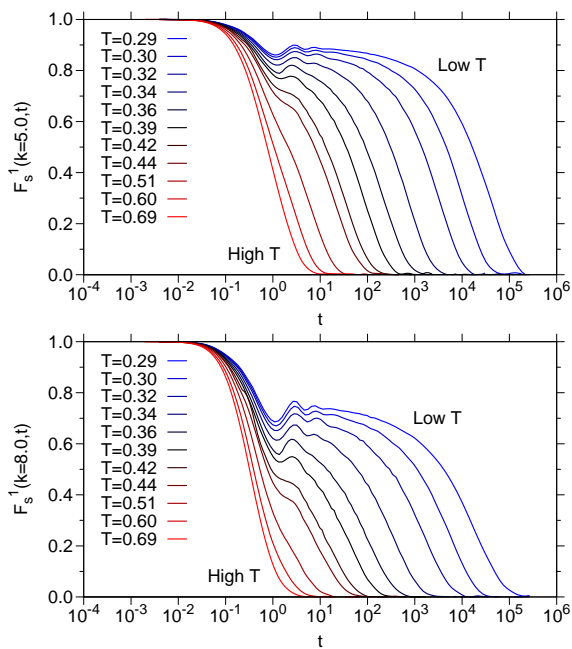}
\end{center}
\caption{\label{fig:fskt} Intermediate scattering functions $F_s(k,t)$
  (self part) at $\rho=1.655$ for wave-vector $k=5.0$ (top panel) and
  $k=8.0$ (bottom panel).}
\end{figure}

Our first step in the description of the dynamical properties of the
NTW model consists in the identification of the so-called
``slow-dynamics regime''~\cite{sastry98}. In this temperature regime,
the dynamical properties of glass-forming liquids assume all their
distinct features, including two-step relaxation, dynamic
heterogeneities, etc. To detect the onset of slow-dynamics, we study
the variation with temperature of the incoherent intermediate
scattering functions
\beq\label{eqn:fskt}
F_s^{\alpha}(k,t) =
\frac{1}{N_{\alpha}}\sum_{i=1}^{N_{\alpha}} \langle \exp
            \left\{i \ve{k} \cdot
            [\ve{r}_i(t)-\ve{r}_i(0)]\right\} \rangle
\eeq
where $\alpha=1,2$ is an index of species.  The $t$-dependence of
$F_s^1(k,t)$ is shown in figure~\ref{fig:fskt} for temperatures in the
range $0.29\leq T\leq 1.50$ at two different wave-numbers: $k =
5.0$, close to the pre-peak in the static structure factors (upper
panel) and $k = 8.0$ (lower panel). Two-step relaxation develops
around $T_O\approx 0.50$, which we take as the onset temperature of
the slow-dynamics regime. Distinct damped oscillations are observed in
$F_s^\alpha(k,t)$ on the time scale of the early $\beta$-relaxation,
i.e., on approaching the plateau. In larger samples (not shown here),
the amplitude of these oscillations is slightly smaller---a well-known
finite-size effect in model network
liquids~\cite{horbach96a,guillot97a}.

\begin{figure}[tb]
\begin{center}
\includegraphics*[width=\twofig]{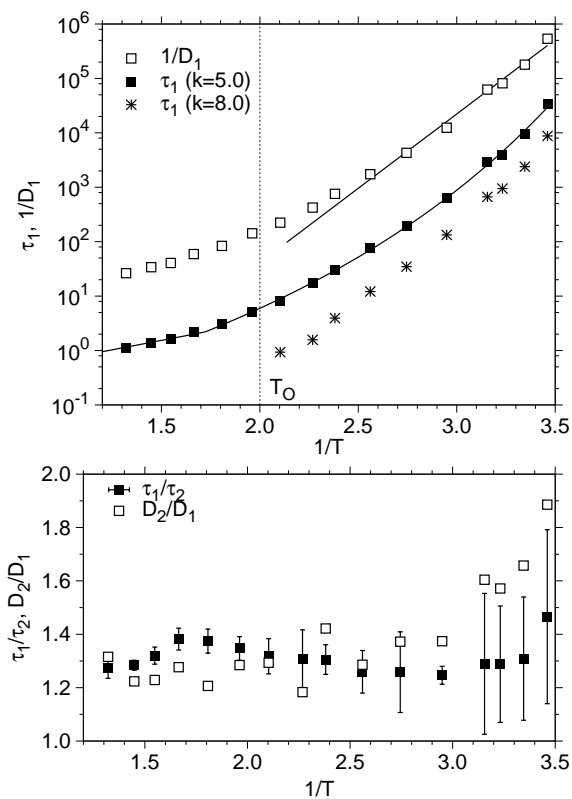}
\end{center}
\caption{\label{fig:angell} Top panel: Angell plot for structural
  relaxation times $\tau_1(k=5.0)$ (filled squares),
  $\tau_1(k=8.0)$ (stars), and inverse diffusion coefficients
  $1/D_1$ (empty squares). All quantities refer to particles of species 1. The
  modified VFT fit for $\tau_1(k=5.0)$ [equation~\eref{eqn:arrvft}] and the
  Arrhenius fit for $1/D_1$ are also shown as solid lines. The dotted
  vertical line marks the onset of the slow dynamics regime. Lower
  panel: Ratios $\tau_1(k=5.0)/\tau_2(k=5.0)$ (filled squares) and $D_2/D_1$ (empty
  squares) as a function of $1/T$.}
\end{figure}

We now analyse the $T$-dependence of the structural relaxation times
extracted from the intermediate scattering functions. Wave-number
dependent relaxation times, $\tau_{\alpha}(k)$, for species $\alpha$
are defined by the condition
$F_s^{\alpha}(k,\tau_{\alpha}(k))=1/e$. In the Angell plot in
Fig.~\ref{fig:angell} we focus on the $T$-dependence of $\tau_1$. We
focus here on the case $\tau\equiv\tau_1(k=5.0)$. To fit the
$T$-dependence of the relaxation times we use the following modified
Vogel-Fulcher equation, previously employed in our study of LJ
mixtures~\cite{coslovich07a},
\beq\label{eqn:arrvft}
\tau(T) = \left\{
\begin{array}{ll}
  \tau_{\infty}     \exp\left[E_{\infty}/T\right] & T>T^*   \\
  \tau_{\infty}^{'} \exp\left[\case{1}{K(T/T_0-1)}\right] & T<T^* \\
\end{array}
\right.
\eeq
where
\beq \tau_{\infty}^{'}=\tau_{\infty} \exp\left[E_{\infty}/T^{*} -
  \case{1}{K(T^{*}/T_0-1)}\right]
\eeq

Equation~\eref{eqn:arrvft} describes the crossover from Arrhenius to
Vogel-Fulcher $T$-dependence of $\tau$, ensuring continuity at
$T=T^*$. Its use in the case of a network glass-formers is justified
by the observation that network liquids display a mild super-Arrhenius
behaviour around and slightly below $T_O$. As we can see from
figure~\ref{fig:angell}, equation~\eref{eqn:arrvft} fits rather well
$\tau(T)$ over about 5 decades. Note that the degree of
super-Arrhenius behaviour in $\tau(T)$ is indeed rather modest and more
visible at wave-numbers corresponding to the first sharp diffraction
peak ($k=5.0$).

Also included in figure~\ref{fig:angell} are the partial diffusion
coefficients $D_1(T)$ obtained from the usual Einstein relation. To
describe the $T$-dependence of the diffusion coefficients, we simply
used the Arrhenius law,
$D_\alpha=D_{\infty}^{\alpha}\exp(E_{\alpha}/T)$. By fitting the data
at low temperature ($T<0.4$), we obtain activation energies
$E_1=6.3\approx 3.8\textrm{eV}$ and $E_2=5.9\approx 3.6 \textrm{eV}$,
which are in reasonable agreement with those obtained in the case of
BKS silica for silicon and oxygen atoms,
respectively~\cite{horbach96a}. The difference in the diffusion
coefficients between the two species is analysed in the lower panel of
figure~\ref{fig:angell}, where the ratio $D_1/D_2$ is shown as a
function of $1/T$. This ratio becomes $\sim 2$ at the lowest
temperatures. A smaller separation of time scales is found when
inspecting the ratio $\tau_2/\tau_1$.

\begin{figure}[tb]
\begin{center}
\includegraphics*[width=0.70\twofig]{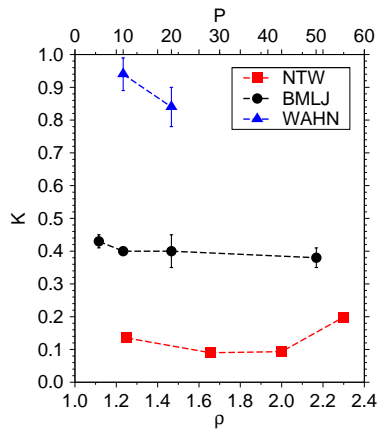} 
\end{center}
\caption{\label{fig:fragility} Fragility index $K$ obtained for NTW
  model and two representative LJ mixtures as a function of $\rho$
  (NTW, lower axis) and $P$ (LJ, upper axis).
}
\end{figure}

The fragility index $K$ of the NTW model obtained from fits to
equation~\eref{eqn:arrvft} is shown in figure~\ref{fig:fragility} as a
function of $\rho$. The system is slightly stronger (smaller $K$) at
densities close to the experimental density of silica. In this range
of density ($1.5<\rho<1.8$), tetrahedral local order becomes nearly
ideal at low $T$. Hence, our results provide support for the link
between structure and dynamic behaviour in network liquids
demonstrated in~\cite{demichele06a} for patchy colloidal
particles. Interestingly, the fragility of the NTW model seems to
increase outside the density range mentioned above \textit{both} at
high \textit{and} low density. Further investigations at low density
would be required to clarify the nature of this behaviour and its
possible connection with a reversibility
window~\cite{phillips81a,thorpe83a}, whose existence for silica has
been suggested by recent work~\cite{trachenko03a}.

The use of a common functional form to describe $\tau=\tau(T)$ of NTW
and LJ models allows a direct comparison of their Angell's
fragility. To this end, we also included in figure~\ref{fig:fragility}
the values of $K$ obtained in~\cite{coslovich07a} for the mixture of
Kob and Andersen (BMLJ~\cite{ka1}) and the mixture of Wahnstr{\"o}m
(WAHN~\cite{wahnstrom}). Clearly, both LJ mixtures have larger
fragility indexes. Furthermore, the fragility index of NTW, $K=0.09$,
obtained at $\rho=1.655$ is lower by around a factor of 3 than the
lowest values found for LJ mixtures ($K=0.24$ for
AMLJ-0.60~\cite{coslovich07a}). The fact that the fragility index was
obtained at constant density for the NTW and constant pressure for the
LJ mixtures does not affect substantially our conclusions. Moreover,
even when considering the variation of $K$ with $\rho$, the largest
fragility index of the NTW model ($K=0.20$ at $\rho=2.300$) is
comparable to the lowest ones found in LJ
mixtures~\cite{coslovich07a}. Thus, despite the presence of
super-Arrhenius behaviour around the onset of slow dynamics, our
network glass-former is stronger than all LJ mixtures studied
in~\cite{coslovich07a}, a fact which fits naturally in the Angell
classification scheme. Moreover, our analysis does not exclude the
occurrence, at low temperatures, of a fragile-to-strong transition---a
scenario which has recently found support on the basis of an energy
landscape approach~\cite{saksaengwijit04a}.

\subsection{Dynamic heterogeneity}

Figure~\ref{fig:fragility} shows that NTW, BMLJ, and WAHN may be
considered as models of strong, intermediate, and fragile
glass-formers, respectively. This offers the opportunity to investigate
the main trends of variations of the dynamics in liquids with
different fragility. In this section, we focus on the degree of
dynamic heterogeneity of the above mentioned models.

As a simple measure of the degree of heterogeneity of the dynamics we will use
the non-Gaussian parameter
\beq\label{eqn:alpha2}
\alpha_2(t)=\frac{3\lav r^4(t)\rav}{5{\lav r^2(t)\rav}^2}-1
\eeq
which measures the deviation of the distribution of particles' displacements
$r(t)$ from a Gaussian distribution. Upon cooling the liquid below $T_O$, in
fact, the distribution of particles' displacements deviates progressively from
a Gaussian and the amplitude of $\alpha_2$ increases. Within the late
$\beta$-relaxation time scale, the non-Gaussian parameter of typical
glass-forming liquids displays a broad peak, whose the position $t^*$ and the
height $\alpha^*$ increase by decreasing temperature.
The trends of variation of the maximum of the non-Gaussian parameter
$\alpha_2^*$ have been found to follow qualitatively the behaviour of more
refined dynamic indicators~\cite{vogel04a}, such as those obtained from four-point correlations
functions~\cite{berthier_spontaneous_2007,berthier_spontaneous_2007-1}.

\begin{figure}
\begin{center}
\includegraphics*[width=\twofig]{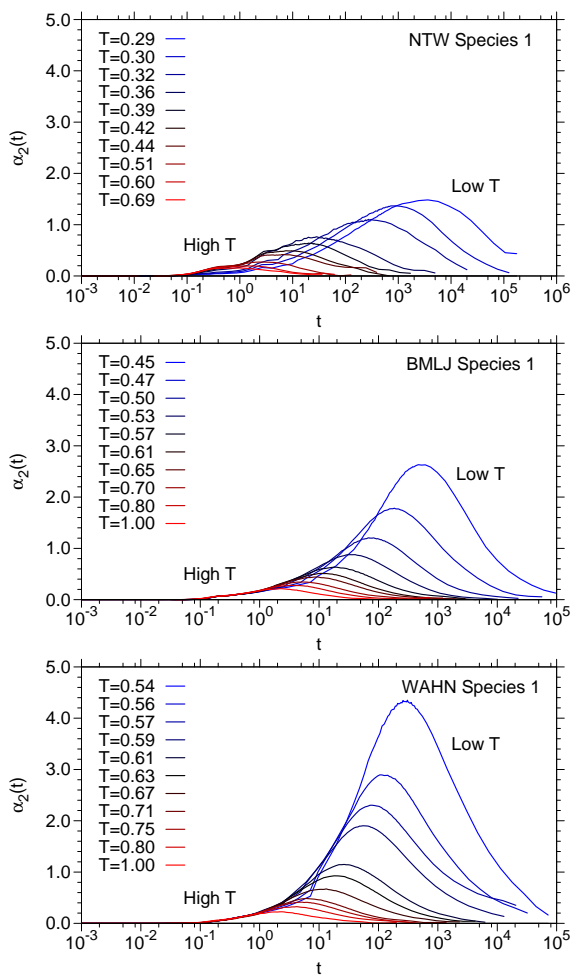}
\end{center}
\caption{\label{fig:a2:1:rho} Non-Gaussian parameter $\alpha_2(t)$ for
  particles of species~1 along isochoric quenches for NTW at
  $\rho=1.655$ (top), BMLJ at $\rho=1.2$ (middle), and WAHN at
  $\rho=1.297$ (bottom).}
\end{figure}

\begin{figure}
\begin{center}
\includegraphics*[width=\twofig]{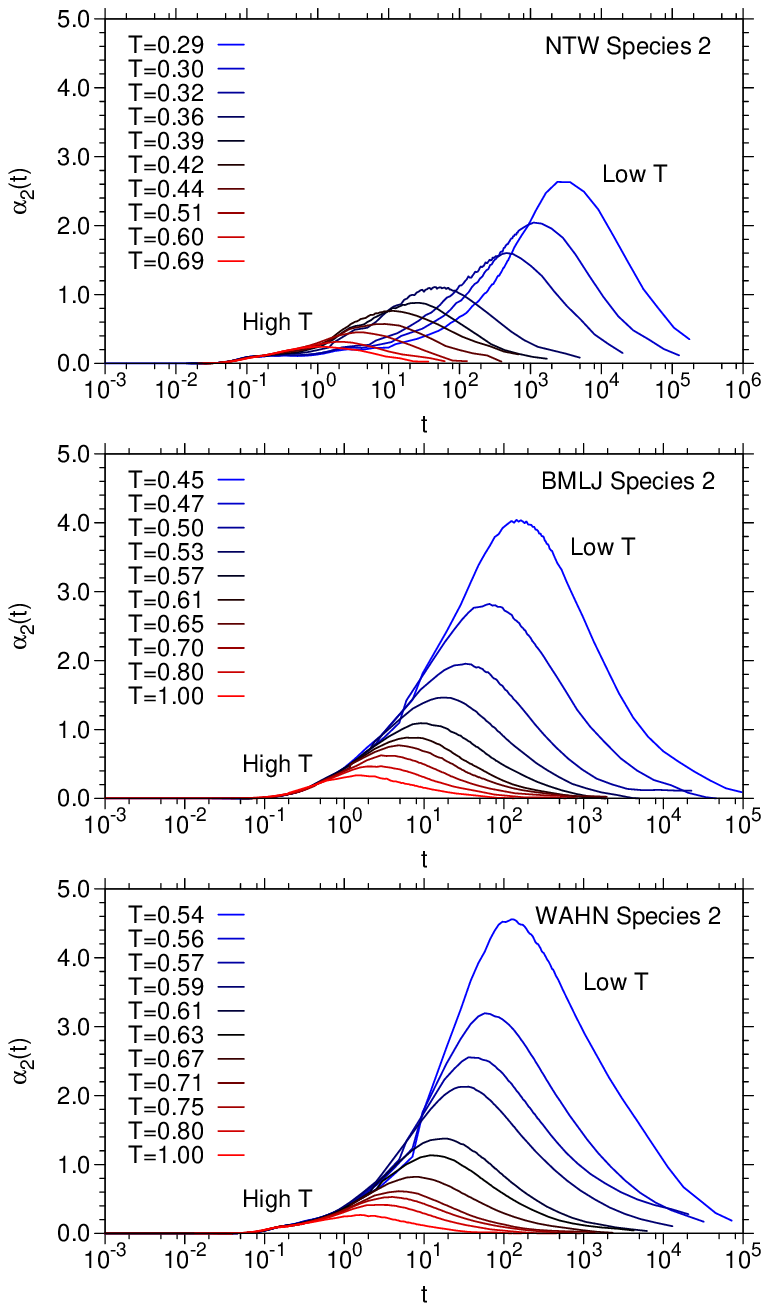}
\end{center}
\caption{\label{fig:a2:2:rho} Same as Fig.~\ref{fig:a2:1:rho} but for
  particles of species 2.  }
\end{figure}

We computed the non-Gaussian parameter $\alpha_2(t)$ in
equation~\eref{eqn:alpha2} separately for species~1 and 2. The results
obtained for NTW, BMLJ, WAHN models along isochoric quenches are shown
in Fig.~\ref{fig:a2:1:rho} and \ref{fig:a2:2:rho} for species~1 and 2,
respectively. The degree of dynamic heterogeneity, as measured from
the height $\alpha_2^*$ of the peak, is least pronounced in the case
of the NTW model close to the ideal density for tetrahedral network
structure. This is consistent with the analysis of
Vogel~\etal~\cite{vogel04a}, who found in fact that BKS silica had a
lower degree of dynamic heterogeneity than other simple glass-formers,
including the BMLJ model. Our results thus indicate a
broad correlation between fragility and the degree of dynamic
heterogeneity in glass-forming liquids. In particular, within the
slow-dynamics regime, both the local structure and the local dynamics
appear more homogeneous in network than in close-packed glass-formers.

\subsection{Local rearrangements}

We now turn to a closer inspection of the nature of local
rearrangements in our model network glass-former. In particular, we
want to identify the structural modifications that accompany
relaxation events. This is motivated by the current interest in
investigating the link between structure and dynamics in glass-forming
liquids~\cite{widmercooper04,widmercooper05}. We will contrast the
results for the NTW model to those previously obtained for LJ
systems~\cite{coslovich06,coslovich07a,coslovich07b}.

In a first attempt to characterize the local dynamics of the NTW model and to
establish a connection with its local structural properties, we computed the
``propensity of motion'' of particles, according to the definition of
Widmer-Cooper~\etal~\cite{widmercooper04}. In this approach, time-dependent
atomic displacements $\Delta r(t)$, relative to a reference configuration, are
averaged over several trajectories generated by independent initial sets of
velocities (``iso-configurational ensemble''). The resulting spatial
distribution of average displacements $\langle \Delta r(t)\rangle_{ic}$, where
$\langle\dots\rangle_{ic}$ denotes an average in the iso-configurational
ensemble, is thus strictly associated to the initial configuration, and can be
used, in principle, to identify the local structural features responsible for
relaxation events.

\begin{figure*}[tb]
\begin{center}
\includegraphics*[width=\onefig]{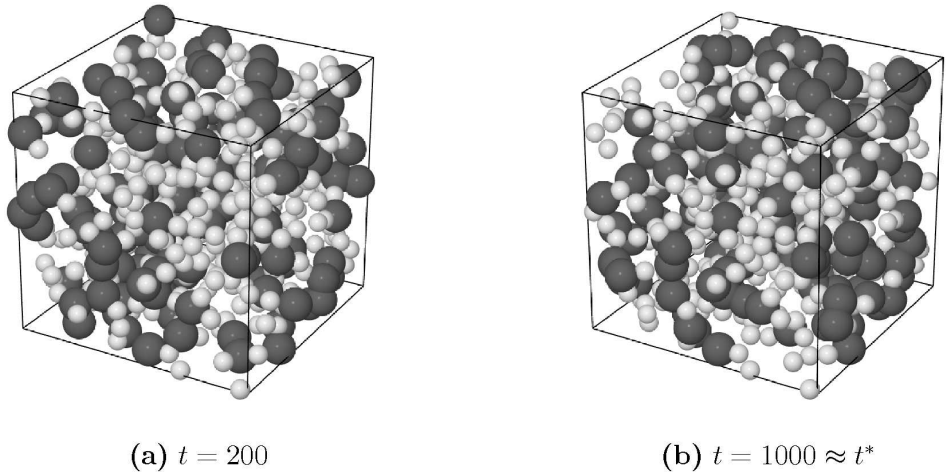}
\end{center}
\caption{\label{fig:propensity} Snapshots of particles having large
  propensity of motion $\langle \Delta r(t) \rangle_{ic}$ for $t=200$
  (left panel) and $t=1000\approx t^*$ (right panel). For both times,
  the 30\% most mobile particles of either species are shown as large
  dark spheres, irrespectively of chemical species. The remaining
  particles are shown as small light spheres.}
\end{figure*}

The spatial distribution of the particles with large propensity of
motion for a representative configuration sampled at $T=0.31$ is shown
in figure~\ref{fig:propensity} for $t=200<t^*$ (left panel) and for
$t=1000\approx t^*$ (right panel). Mobile particles, depicted in
figure~\ref{fig:propensity} as large dark spheres, are identified as
the ones having the 30\% largest propensities of motion among those of
the same chemical species.
The overall picture does not change upon small variation of the
fraction of particles displayed. There is no substantial clustering of
mobile particles on either time scales. Some clustering is observed at
$t=1000$ but the size of the clusters remains rather modest (less than
$\sim$ 10 neighbouring particles). This is strikingly different from the
results obtained in LJ systems~\cite{coslovich06} within the
slow-dynamics regime. In LJ systems, a pronounced clustering of
particles with large propensity of motion has been observed for times
on the order of the late $\beta$-relaxation ($t\approx t^*$). Our
results confirm that the degree of dynamic heterogeneity of network
liquids is much less pronounced than in close-packed LJ systems, and
show that the origin of the weak dynamic heterogeneity observed within
the $\alpha$-relaxation time scale is essentially kinetic, rather than
structural.

\begin{figure*}[tb]
\begin{center}
\includegraphics*[width=\twofig]{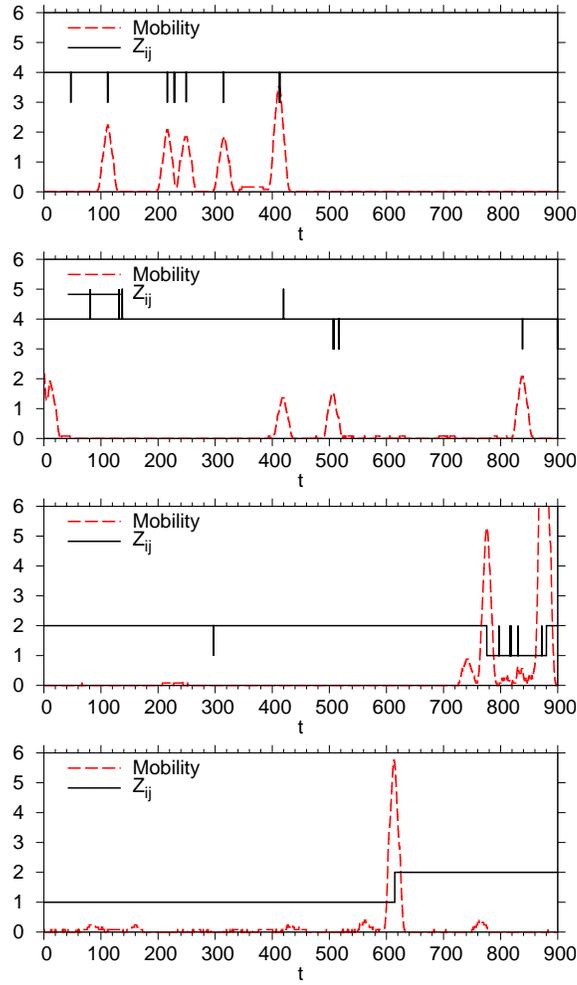}
\end{center}
\caption{\label{fig:mobility} Instantaneous mobility $\mu_i(t)$ in
  arbitrary units (dashed line) for two representative particles of
  species 1 (two upper panels) and species 2 (two lower panels) at
  $T=0.29$. Also shown are the instantaneous coordination numbers
  $Z_{12}(t)$ and $Z_{21}(t)$ for particles of species 1 and 2,
  respectively (solid lines).}
\end{figure*}

The results above do not imply, however, that there is no link at all
between local structure and dynamics in network liquids. Such a link
is more subtle and requires a different type of investigation. The
breaking and reformation of bonds involved in typical relaxation
events~\cite{horbach96a} occurs, in fact, on a very short time scale
and the averaging introduced by the iso-configurational ensemble
washes out this information. To overcome this problem we calculated,
following Ladadwa and Teichler~\cite{ladadwa06}, the instantaneous
mobility of particles from a smoothed atomic trajectory
\begin{equation}
\bar{\ve{r}}_i(t)=\int_{-\infty}^\infty \ve{r}_i(t') \phi(t',t)  dt'
\end{equation}
where $\phi(t',t)$ is a smoothing function normalized to 1. Rather than a
Gaussian~\cite{ladadwa06}, we used a simple window smoothing function of length
$2\Delta t$, which equals $1/(2\Delta t)$ for $t-\Delta
t<t'<t-\Delta t$ and 0 otherwise. From the smoothed trajectories,
we computed the instantaneous atomic mobility~\cite{ladadwa06}
\begin{eqnarray*}
\mu_i(t)^2 &= &\int_{-\infty}^\infty
[\bar{\ve{r}}_i(t)-\bar{\ve{r}}_i(t')]^2 \phi(t',t)  dt' \\ &= & 
\frac{1}{2\Delta t}\int_{t-\Delta t}^{t+\Delta t}
[\bar{\ve{r}}_i(t)-\bar{\ve{r}}_i(t')]^2  dt'
\end{eqnarray*}
using $\Delta t=10$. The time dependence of $\mu_i(t)$ is shown in
figure~\ref{fig:mobility} for representative particles of species 1
and 2 at $T=0.29$. At this low temperature, atomic mobilities show
intermittent behaviour, with long periods of inactivity (vibrations)
followed by displacements occurring on a very short time scale. To
establish the connection with the changes in the local structure, we
also plot, for the same time interval, the instantaneous coordination
number $Z(t)$. A clear correlation between intermittent dynamical
events and bond breaking and reformation processes is observed. In
particular, defective local environments ($Z_{12}\neq 4$ and
$Z_{21}\neq 2$), either created instantaneously by thermal
fluctuations or associated to long-lived defective configurations, are
closely associated to dynamical events. Interestingly, this shows that
in network liquids the link between structure and dynamics can be
understood at a single-particle level. Such a link has been
demonstrated in LJ systems only at a coarse-grained spatial
level~\cite{berthier07b}.

\begin{figure*}[tb]
\begin{center}
\includegraphics*[width=\onefig]{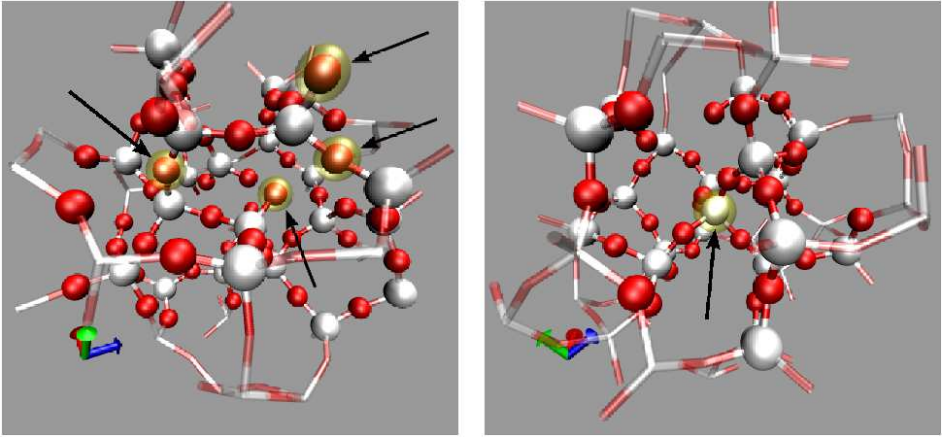}
\end{center}
\caption{\label{fig:movies} Animations of representative elementary
  dynamical events at $T=0.29$ (see
  \url{http://www-dft.ts.infn.it/media/gp/1/} for the corresponding MPG
  files, realized with VMD~\cite{HUMP96}). White large spheres and
  small red spheres correspond to particles of species 1 and 2,
  respectively. The particles involved in the elementary dynamical
  events are surrounded by a yellow halo. Atomic positions have been
  averaged over a time window of 4.2 reduced time units (700 time
  steps) to remove thermal motion and help visualization. Left panel:
  four small particles (indicated by arrows in the figure) in the
  central-upper part of the figure perform a correlated, rotational
  motion in neighbouring tetrahedra (file movie1.mpg; size: 1.4
  Mb). Right panel: the central particle (indicated by an arrow)
  explores its low density environment with a sequence of two jumps,
  associated to bond breaking and reformation (file: movie2.mpg; size:
  1.7 Mb). }
\end{figure*}

Finally, we describe qualitatively the typical local rearrangements
observed at low temperature in the NTW model. By inspection of
animated atomic trajectories, we identified two typical relaxation
processes, closely related to the ones occurring in BKS silica. See
figure~\ref{fig:movies} and supplementary materials
(available at \url{http://www-dft.ts.infn.it/media/gp/1/}) for two representative
events. A first class of rearrangements involves correlated rotations
of tetrahedra formed by small particles around nearly immobile large
particles. The overall process resembles the ``rotational period''
recently described by Heuer and coworkers~\cite{saksaengwijit07a}, in
which oxygens perform permutation of the tetrahedral positions around
a fixed silicon atom. A second class of rearrangements is closely
related to the ones described by Horbach and
Kob~\cite{horbach96a}. One large particle jumps out from one of the
faces of the tetrahedron surrounding it and attaches itself to an
under-coordinated ($Z_{21}=1$) small particle. At the same time, a
slight recoil movement of the small particles forming the involved
tetrahedron is observed, together with the formation of a new dangling
bond. Contrary to rotational rearrangements, which often involve a few
neighbouring tetrahedra, this second class of elementary dynamical
events is strongly localized around the involved tetrahedron. On
longer time scales, however, sequences of independent events are also
observed (see right panel of figure~\ref{fig:movies}).

\section{Stationary points and unstable modes}\label{sec:pes}

Summarising our previous analysis, two key features characterize the
dynamical behaviour of our model network liquid: strong behaviour in
the Angell's classification scheme and a significant homogeneity of
atomic displacements within the late $\beta$-relaxation time scale. In
this section, we rationalize these features in terms of the properties
of the Potential Energy Surface (PES). In particular, we first provide
an estimate of the average energy barriers in the PES of the NTW
model, and then analyse the localization properties and real-space
structure of the unstable modes associated to stationary points of the
PES.

\subsection{Energy barriers}

The nature and distribution of barriers connecting stationary points
of the PES have been long recognized as key aspects for understanding
the dynamics of fragile and strong
liquids~\cite{stillinger95,debenedetti01a}. Recently, sophisticated
analysis of transitions between meta-basins for model glass-forming
liquids~\cite{heuer_exploringpotential_2008} have provided even
further quantitative evidence of the importance of the PES. In this
section, we employ a simple definition of average energy
barriers~\cite{cavagna01c,coslovich07b} to quantify the roughness of
the potential energy surface of the NTW model.

Our analysis of the PES is based on the procedure described
in~\cite{coslovich07b}. For each state point, we perform minimizations
of the mean square total force $W$ to locate the closest stationary
points along the dynamical trajectory. Typically, between 100 and 400
configurations per state point are considered as starting points for
$W$-minimizations. It is well known that $W$-minimizations often
locate points with a low value of $W$ ($W \approx 10^{-2}\div
10^{-4}$) that are not true stationary points. These points, usually
called quasi-saddles, contain nonetheless relevant information about
the dynamics~\cite{angelani02,angelani03}. In the following, we will
include these points in our analysis, without further distinction
between true stationary points and quasi-saddles. Having located the
stationary points, we diagonalize the Hessian matrix of the potential
energy surface and thus obtain a set of $3N$ eigenfrequencies
$\omega_\alpha$ and eigenvectors $\{\ve{e}_i^\alpha\}$. The unstable
eigenvectors ($\omega_\alpha^2<0$) are of particular interest for our
discussion, because they are more directly related to the dynamical
behaviour of the system~\cite{coslovich06}. As in previous
work~\cite{coslovich07b}, we will report the imaginary branch of the
frequency spectrum along the real negative axis.

\begin{figure}[tb]
\begin{center}
\includegraphics*[width=\twofig]{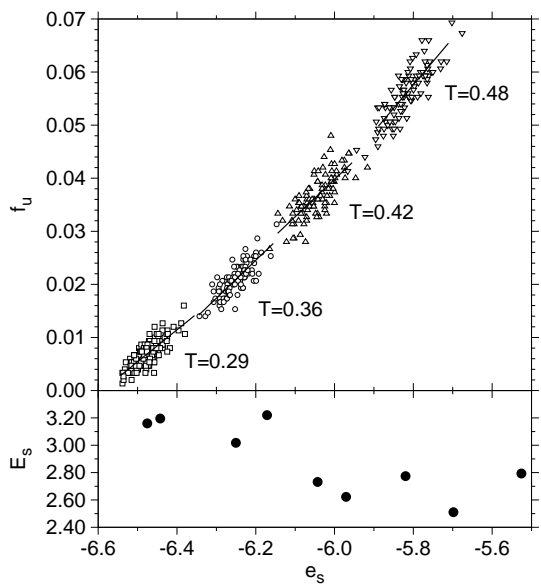}
\end{center}
\caption{\label{fig:barriers} Upper plot: fraction
  of unstable modes $f_u$ as a function of the energy $e_s$ of
  individual stationary points sampled at the indicated
  temperatures. Also included are linear fits, from which the average
  energy barriers $E_s$ are obtained. Lower panel: average energy
  barriers $E_s$ as a function of the average energy of stationary
  points $e_s=e_s(T)$ for various $T$.  }
\end{figure}

To estimate the average potential energy barriers we follow the
definition given by Cavagna~\cite{cavagna01c}
\begin{equation}
E_s = \frac{1}{3} \der{e_s}{f_u}
\end{equation}
where $e_s=e_s(f_u)$ is the average energy of stationary points having
a fraction of unstable modes $f_u=n_u/3N$. As in~\cite{coslovich07b},
we evaluate $E_s$ from the slope obtained by linear regression of
$e_s$ vs. $f_u$ of individual stationary points sampled at temperature
$T$. The procedure is illustrated in the upper panel of
figure~\ref{fig:barriers}, where $f_u$ is shown as a function of $e_s$
for stationary points sampled at selected temperatures ($T<T_O$). The
energy barriers obtained for individual state points are collected in
the bottom panel as a function of the average energy of stationary
points $e_s=e_s(T)$.

From the comparison of the results above with those obtained
in~\cite{coslovich07b} for LJ mixtures, we draw two main conclusions,
which highlight the peculiarity of the network liquids: (i) energy
barriers in the NTW model are large compared to typical thermal
energies already for $T\approx T_O$, and (ii) their increase is very
weak (less than 20\%) with decreasing temperature below $T_O$. At
least at a qualitative level, (ii) confirms our
conjecture~\cite{coslovich07b} that the fragility of a glass-forming
liquid is related to the increase of average energy barriers $E_s$
upon cooling below $T_O$. It also supports the overall picture that
the energy landscape of network liquids has a uniformly rough
structure, with barriers whose average amplitude is nearly independent
of the energy level. Organization of stationary points into meta-basin
structures, while present even in network liquids~\cite{heuer}, should
be of much more limited extent than in the more fragile, close-packed
glass-formers.

\subsection{Localization properties}

We now investigate the localization properties and the real-space structure of the
unstable eigenvectors of the stationary points sampled in the
slow-dynamics regime. In particular, we aim at explaining the more
``homogeneous'' character of atomic displacements observed in NTW, compared to
the more fragile LJ mixtures.

\begin{figure}[tb]
\begin{center}
\includegraphics*[width=\twofig]{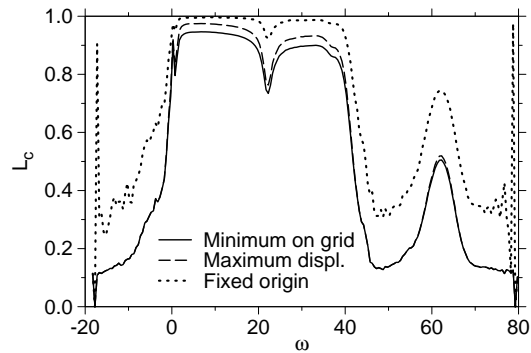}
\end{center}
\caption{\label{fig:lc} Average gyration radius $L_c$ of modes of
  frequency $\omega$ calculated at $\rho=1.655$ and $T=0.31$ using
  different procedures (see text for definitions): straightforward
  calculation using equation~\eref{eqn:lc} (dotted line), method (i)
  (dashed line), and method (ii) (solid line).}
\end{figure}

One possible measure of the degree of mode localization, used in a
number of previous
investigations~\cite{marinov_model_1997,caprion,ispas_vibrational_2005},
is provided by the gyration radius
\begin{equation}\label{eqn:lc}
L_c^\alpha = \sum_{i=1}^N {|e_i^\alpha|}^2|\ve{r}_i-\ve{r}_{c}|^2
\end{equation}
where $r_c=\sum_{i=1}^N \ve{r}_i {|e_i^\alpha|}^2$ is the ``centre of
mass'' of the mode. Extended modes should have $L_c\approx 1.0$, while
$L_c<1.0$ for localized modes. It turns out, however, that the
definition in equation~\eref{eqn:lc} is inappropriate for systems with
periodic boundary conditions. The centre of mass of the mode, in fact,
is not well defined in a periodic system, and the value of $L_c$ thus
depends on the choice of the origin of the frame of coordinates. As it
will be clear in the following, this shortcoming is particularly
evident in the case of strongly localized modes. To overcome this
problem, we employ two alternative definitions of the gyration radius,
obtained by redefining the origin of the system coordinates: (i) the
position of the particle that has the largest displacement on mode
$\alpha$ is used as the origin of the system coordinates for the
calculation of $L_c^\alpha$; (ii) the gyration radius is determined by
minimization of $L_c^\alpha$ over all possible origins of the system
coordinates chosen on a grid of points subdividing the cubic cell.

\begin{figure}[tb]
\begin{center}
\includegraphics*[width=\twofig]{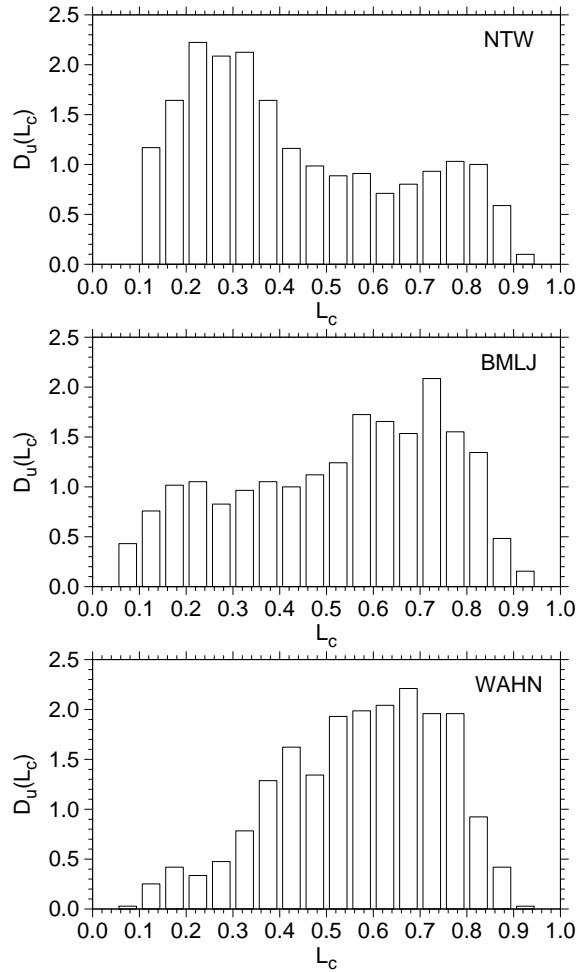}
\end{center}
\caption{\label{fig:lc_unstable} Distribution of $L_c$ for unstable
  modes sampled in the slow-dynamics regime: in NTW at $\rho=1.655$,
  $T=0.31$ (upper panel), in BMLJ at $\rho=1.2$, $T=0.45$ (middle
  panel), in WAHN at $\rho=1.297$, $T=0.54$ (lower panel).  }
\end{figure}

In figure~\ref{fig:lc} we show the average gyration radius $L_c$ of
modes with frequency $\omega$ obtained for NTW at $T=0.31$, using the
original definition and the two alternatives (i) and (ii). The
original definition substantially overestimates the extension of
strongly localized modes, while discrepancies are somehow less
pronounced for extended modes. For extended modes, smaller
discrepancies are apparent between methods (i) and (ii). In the
following, we will employ definition (ii). Only minor quantitative
differences in the following analysis appear when using definition
(i).

We now focus on the unstable modes, which have been found to contain
direct information on the dynamics of a glass-forming
liquid~\cite{coslovich06}. In figure~\ref{fig:lc_unstable} we show the
distribution of $L_c$ for the NTW model at $T=0.29$. For comparison,
we also show analogous distributions obtained for BMLJ and WAHN at
temperatures deep in the slow-dynamics regime. The distribution of
$L_c$ for NTW is bimodal, with an excess of localized unstable modes
having $L_c\sim 0.2$. A similar, yet much less pronounced, excess peak
at low $L_c$ is observed in the distribution for BMLJ, while no such
feature is found for the very fragile WAHN model. Justified by the
fact that the minimum of the distribution of $L_c$ for NTW is located
around 0.5, we define a mode localized (extended) if $L_c$ is smaller
(larger) than 0.5. The precise location of this cutoff is irrelevant
for the discussion below.

\begin{figure}[tb]
\begin{center}
\includegraphics*[width=\twofig]{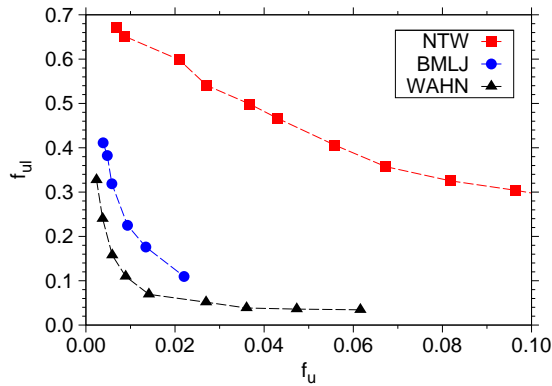}
\end{center}
\caption{\label{fig:lc_fraction} Fraction of localized unstable modes
  $f_{ul}$ in stationary points as a function of the total fraction
  unstable modes $f_u$ in NTW at $\rho=1.655$ (squares), BMLJ at
  $\rho=1.2$ (circles), and WAHN at $\rho=1.297$ (triangles).}
\end{figure}

Information about the relevance of localized unstable modes in
different glass-formers is conveniently represented by the plot in
figure~\ref{fig:lc_fraction}, where the fraction of localized unstable
modes $f_{ul}$ is shown as a function of $f_u$.
The NTW model has a substantial fraction of localized unstable modes,
independent of the energy level in the PES. $f_{ul}$ only weakly
increases by lowering $f_u$, i.e., as the systems explores lower
regions of the energy landscape. In contrast, in the very fragile WAHN
$f_{ul}$ is very small and increases only on approaching the bottom of
the energy landscape. The BMLJ displays an intermediate trend, since a
larger fraction of localized unstable modes is present. These
localized modes of the BMLJ mixture correspond to eigenvectors where a
small particle and few other neighbours have large displacements (see
figure 13 in~\cite{coslovich07b}). It is remarkable that the trend
observed in the localization of unstable modes follows the different
dynamic character (strong, intermediate, fragile) of the models
studied (see also~\cite{jagla01}). Moreover, our results provide a
simple explanation of the weak degree of dynamic heterogeneity in
network liquids in terms of an excess of localized, uncooperative
unstable modes, which are absent, or at least rarer, in the more
fragile LJ mixtures.

\begin{figure*}[tb]
\begin{center}
\includegraphics*[width=\onefig]{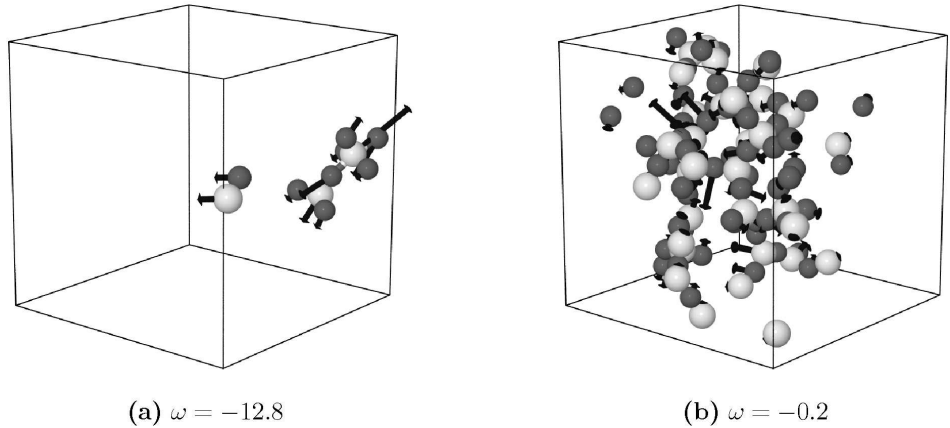}
\end{center}
\caption{\label{fig:modes} Typical extension of localized (a) and
  extended (b) unstable modes of a stationary point sampled at
  $T=0.29$. Only particles with displacement $|e_i^\alpha|$ larger
  than $0.04$ are shown, and the eigenvectors are scaled
  logarithmically. Particles of species 1 and 2 are shown as white
  large spheres and small dark spheres, respectively.}
\end{figure*}

Finally, we describe the nature of the rearrangements associated to
localized and extended unstable modes of the NTW model. The typical
extensions of localized and extended unstable modes are depicted in
figure~\ref{fig:modes}. 
From inspection of the real-space structure of the atomic
displacements we found that the unstable modes reproduce the two classes
of local relaxation processes described in
section~\ref{sec:results}. Extended unstable modes usually involve
coupled rotations of tetrahedra and have a marked collective
nature. 
These modes are thus good candidates for explaining the rotational
motions described in section~\ref{sec:results}. Similar unstable modes
have been found in the unstable branch of the instantaneous normal
mode spectrum of BKS silica~\cite{bembenek_instantaneous_2001}. On the
other hand, localized unstable modes correspond rather well to the
second class of local rearrangements described in
section~\ref{sec:results}. The mode depicted in left panel of
figure~\ref{fig:modes} shows the passage of a large particle,
initially at the centre of a tetrahedron, through the face of the
tetrahedron, while a neighbouring under-coordinated small particle
moves in the opposite direction to create a bond. In the intermediate
stage along the reaction coordinate of the mode the large particle is
5-fold coordinated. Our results indicate that both types of elementary
dynamical events should be taken into account for a complete
description of the dynamics in network liquids.  Approaches that focus
only on soft {\em stable}
modes~\cite{brito_wyart_2007,widmer-cooper_perry_harrowell_reichman_2008}
may not be able to capture the localized nature of the dynamics in
network liquids. In these systems, in fact, the low frequency portion
of the VDOS encompasses collective modes that typically involve
coupled rotations of
tetrahedra~\cite{buchenau_zhou_nucker_gilroy_phillips_1988,taraskin_nature_1997}.

\section{Conclusions}\label{sec:conclusions}

In this work we have extended our previous
analysis~\cite{coslovich07a,coslovich07b} on the glass transition of
fragile Lennard-Jones mixtures by introducing a new model of
tetrahedral network glass-former based on short-ranged, spherical
interactions. Remarkably, these simple models of liquids, all based on
pair potentials of the Lennard-Jones type, are able to reproduce
qualitatively a wide spectrum of dynamic behaviours, thus allowing
extensive and detailed investigations of the glass transition
phenomenon.

Notwithstanding the problem of crystallization, which may occur in
binary mixtures during longer
simulations~\cite{pedersen_crystallization_2007,valdes_mixing_2009}
but is not observed in our samples, we have found that the fragile
vs. strong behaviour of our models can be clearly identified and
rationalized even at relatively high $T$, but below the onset
temperature of slow dynamics. Using an appropriate parameterization of
the $T$-dependence of relaxation times, we have found that the
model network glass-former is stronger at all studied densities than
all previously investigated LJ mixtures. Our results also confirm that
the degree of dynamic heterogeneity is less pronounced in network than
in close-packed glass-formers.

An important aspect of the glass transition concerns the nature of
atomic rearrangements occurring within the $\alpha$-relaxation
time~\cite{appignanesi,appignanesi_reproducibility_2006}. The relation
between dynamical events and the nature of the local structure is of
particular interest~\cite{widmercooper04,widmercooper05}. The analysis
of the propensity of motion of particles within the late
$\beta$-relaxation time scale, combined with a comparative study of
the non-Gaussian parameter in different systems, has revealed a
substantial homogeneity of atomic mobility in the model network
glass-former. However, contrary to the case of LJ mixtures, it is
possible to establish a relation between local structure and dynamics
at the single-particle level by considering individual atomic
trajectories. Periods of high mobility are in fact clearly associated
to sequences of bond breaking and reformation, i.e. variations in the
local structure.

The features described above and the variation of dynamic behaviour in
systems with different fragility can be rationalized well in terms of
the features of the Potential Energy Surface. We have focused on the
properties of the unstable modes of saddles and quasi-saddles sampled
within the slow-dynamics regime. The amplitude of the average energy
barriers $E_s$ in the model network glass-former is always larger than
typical thermal energies below $T_O$ and depends very mildly on the
energy level. This contrasts the findings in the more fragile LJ
mixtures, where $E_s$ rapidly increases upon entering in the
slow-dynamics regime~\cite{coslovich07b}. The localization of the
unstable modes offers direct insight into the elementary dynamic
events leading to relaxation. In general, as the system explores lower
and lower regions of the energy landscape, the unstable modes soften
and retain a cooperative character. In the NTW model, there is also a
significant fraction of localized unstable modes that persists in the
whole slow-dynamics regime. These localized modes typically describe
bond breaking and reformation, i.e. elementary rearrangements that
characterize the dynamics of the model. On the contrary, close-packed
fragile liquids have a large fraction of extended unstable modes,
which soften and tend to localize only on approaching the bottom of
the landscape. As a result, the dynamics in the latter systems is
inherently more cooperative than in network liquids.

\ack 

D.C. acknowledges financial support from the Austrian Research Fund
(FWF) (Project number: P19890-N16)


\section*{References}

\begin{thebibliography}{10}
\providecommand{\url}[1]{\texttt{#1}}
\providecommand{\urlprefix}{URL }
\providecommand{\eprint}[2][]{\url{#2}}

\bibitem{angell88}
Angell C~A 1988 \emph{J. Phys. Chem. Solids} \textbf{49} 863

\bibitem{saika-voivod_fragile-to-strong_2001}
{Saika-Voivod} I, Poole P~H and Sciortino F 2001 \emph{Nature} \textbf{412} 514

\bibitem{vogel04a}
Vogel M and Glotzer S 2004 \emph{Phys. Rev. E} \textbf{70} 061504

\bibitem{berthier07a}
Berthier L 2007 \emph{Phys. Rev. E} \textbf{76} 011507

\bibitem{garrahan03a}
Garrahan J~P and Chandler D 2003 \emph{Proc. Nat. Acad. Soc.} \textbf{100} 9710

\bibitem{stillinger95}
Stillinger F~H 1995 \emph{Science} \textbf{267} 1935

\bibitem{debenedetti01a}
Debenedetti P~G and Stillinger F~H 2001 \emph{Nature} \textbf{410} 259

\bibitem{vanbeest90}
van Beest B~W~H, Kramer G~J and van Santen R~A 1990 \emph{Phys. Rev. Lett.}
  \textbf{64} 1955

\bibitem{saikavoivod00a}
Saika-Voivod I, Sciortino F and Poole P~H 2000 \emph{Phys. Rev. E} \textbf{63}
  011202

\bibitem{saikavoivod04a}
Saika-Voivod I, Sciortino F, Grande T and Poole P~H 2004 \emph{Phys. Rev. E}
  \textbf{70} 061507

\bibitem{vollmayr96a}
Vollmayr K, Kob W and Binder K 1996 \emph{Phys. Rev. B} \textbf{54} 15808

\bibitem{horbach99a}
Horbach J and Kob W 1999 \emph{Phys. Rev. B} \textbf{60} 3169

\bibitem{horbach01a}
Horbach J and Kob W 2001 \emph{Phys. Rev. E} \textbf{64} 041503

\bibitem{taraskin_nature_1997}
Taraskin S~N and Elliott S~R 1997 \emph{Phys. Rev. B} \textbf{56} 8605

\bibitem{taraskin_anharmonicity_1999}
Taraskin S~N and Elliott S~R 1999 \emph{Phys. Rev. B} \textbf{59} 8572

\bibitem{benoit02a}
Benoit M and Kob W 2002 \emph{Europhys. Lett.} \textbf{60} 269

\bibitem{jund_computer_1999}
Jund P and Jullien R 1999 \emph{Phys. Rev. Lett.} \textbf{83} 2210

\bibitem{la_nave_configuration_2002}
Nave E~L, Stanley H~E and Sciortino F 2002 \emph{Phys. Rev. Lett.} \textbf{88}
  035501

\bibitem{bembenek_instantaneous_2001}
Bembenek S~D and Laird B~B 2001 \emph{J. Chem. Phys.} \textbf{114} 2340

\bibitem{saksaengwijit04a}
Saksaengwijit A, Reinisch J and Heuer A 2004 \emph{Phys. Rev. Lett.}
  \textbf{93} 235701

\bibitem{kerrache05a}
Kerrache A, Teboul V and Monteil A 2006 \emph{Chem. Phys.} \textbf{321} 69

\bibitem{carre07a}
Carr\'e A, Berthier L, Horbach J, Ispas S and Kob W 2007 \emph{J. Chem. Phys.}
  \textbf{127} 114512

\bibitem{ford04a}
Ford M~H, Auerbach S~M and Monson P~A 2004 \emph{J. Chem. Phys.} \textbf{121}
  8415

\bibitem{demichele06a}
{De Michele} C, Tartaglia P and Sciortino F 2006 \emph{J. Chem. Phys.}
  \textbf{125} 204710

\bibitem{zaccarelli_spherical_2007}
Zaccarelli E, Sciortino F and Tartaglia P 2007 \emph{J. Chem. Phys.}
  \textbf{127} 174501

\bibitem{ferrante89a}
Ferrante A and Tosi M~P 1989 \emph{J, Phys.: Condens. Matter} \textbf{1} 1679

\bibitem{coslovich07a}
Coslovich D and Pastore G 2007 \emph{J. Chem. Phys.} \textbf{127} 124504

\bibitem{coslovich07b}
Coslovich D and Pastore G 2007 \emph{J. Chem. Phys.} \textbf{127} 124505

\bibitem{grigera02}
Grigera T, Cavagna A, Giardina I and Parisi G 2002 \emph{Phys. Rev. Lett.}
  \textbf{88} 055502

\bibitem{horbach96a}
Horbach J, Kob W, Binder K and Angell C~A 1996 \emph{Phys. Rev. E} \textbf{54}
  5897(R)

\bibitem{carpenter85a}
Carpenter J~M and Price D~L 1985 \emph{Phys. Rev. Lett.} \textbf{54} 441

\bibitem{sastry98}
Sastry S, Debenedetti P~G and Stillinger F~H 1998 \emph{Nature} \textbf{393}
  554

\bibitem{jin94a}
Jin W, Kalia R~K, Vashishta P and Rino J~P 1994 \emph{Phys. Rev. B} \textbf{50}
  118

\bibitem{wischnewski98a}
Wischnewski A, Buchenau U, Dianoux A~J, Kamitakahara W~A and Zarestky J~L 1998
  \emph{Phys. Rev. B} \textbf{57} 2663

\bibitem{guillot97a}
Guillot B and Guissani Y 1997 \emph{Phys. Rev. Lett.} \textbf{78} 2401

\bibitem{phillips81a}
Phillips J~C 1981 \emph{J. Non-Cryst. Solids} \textbf{43} 37

\bibitem{thorpe83a}
Thorpe M~F 1983 \emph{J. Non-Cryst. Solids} \textbf{57} 355

\bibitem{trachenko03a}
Trachenko K and Dove M~T 2003 \emph{Phys. Rev. B} \textbf{67} 212203

\bibitem{ka1}
Kob W and Andersen H~C 1995 \emph{Phys. Rev. E} \textbf{51} 4626

\bibitem{wahnstrom}
Wahnstr{\"o}m G 1991 \emph{Phys. Rev. A} \textbf{44} 3752

\bibitem{berthier_spontaneous_2007}
Berthier L, Biroli G, Bouchaud J, Kob W, Miyazaki K and Reichman D~R 2007
  \emph{J. Chem. Phys.} \textbf{126} 184503

\bibitem{berthier_spontaneous_2007-1}
Berthier L, Biroli G, Bouchaud J, Kob W, Miyazaki K and Reichman D~R 2007
  \emph{J. Chem. Phys.} \textbf{126} 184504

\bibitem{widmercooper04}
Widmer-Cooper A, Harrowell P and Fynewever H 2004 \emph{Phys. Rev. Lett.}
  \textbf{93} 135701

\bibitem{widmercooper05}
Widmer-Cooper A and Harrowell P 2005 \emph{J. Phys.: Condens. Matter}
  \textbf{17} S4025

\bibitem{coslovich06}
Coslovich D and Pastore G 2006 \emph{Europhys. Lett.} \textbf{75} 784

\bibitem{ladadwa06}
Ladadwa I and Teichler H 2006 \emph{Phys. Rev. E} \textbf{73} 031501

\bibitem{berthier07b}
Berthier L and Jack R~L 2007 \emph{Phys. Rev. E} \textbf{76} 041509

\bibitem{HUMP96}
Humphrey W, Dalke A and Schulten K 1996 \emph{J. Mol. Graphics} \textbf{14} 33

\bibitem{saksaengwijit07a}
Saksaengwijit A and Heuer A 2007 \emph{J. Phys.: Condens. Matter} \textbf{19}
  205143

\bibitem{heuer_exploringpotential_2008}
Heuer A 2008 \emph{J. Phys.: Condens. Matter} \textbf{20} 373101

\bibitem{cavagna01c}
Cavagna A 2001 \emph{Europhys. Lett.} \textbf{53} 490

\bibitem{angelani02}
Angelani L, {Di Leonardo} R, Ruocco G, Scala A and Sciortino F 2002 \emph{J.
  Chem. Phys.} \textbf{116} 10297

\bibitem{angelani03}
Angelani L, Ruocco G, Sampoli M and Sciortino F 2003 \emph{J. Chem. Phys.}
  \textbf{119} 2120

\bibitem{heuer}
Heuer A and Buchner S 2000 \emph{J. Phys,: Condens. Matter} \textbf{12} 6535

\bibitem{marinov_model_1997}
Marinov M and Zotov N 1997 \emph{Phys. Rev. B} \textbf{55} 2938

\bibitem{caprion}
Caprion D and Schober R 2001 \emph{J. Chem. Phys.} \textbf{114} 3236

\bibitem{ispas_vibrational_2005}
Ispas S, Zotov N, Wispelaere S~D and Kob W 2005 \emph{J. Non-Cryst. Solids}
  \textbf{351} 1144

\bibitem{jagla01}
Jagla E~A 2001 \emph{Mol. Phys.} \textbf{99} 753

\bibitem{brito_wyart_2007}
Brito C and Wyart M 2007 \emph{J. Stat. Mech.: Theory Exp.} \textbf{2007}
  L08003

\bibitem{widmer-cooper_perry_harrowell_reichman_2008}
Widmer-Cooper A, Perry H, Harrowell P and Reichman D~R 2008 \emph{Nat. Phys.}
  \textbf{4} 711

\bibitem{buchenau_zhou_nucker_gilroy_phillips_1988}
Buchenau U, Zhou H~M, Nucker N, Gilroy K~S and Phillips W~A 1988 \emph{Phys.
  Rev. Lett.} \textbf{60} 1318

\bibitem{pedersen_crystallization_2007}
Pedersen U~R, Bailey N~P, Dyre J~C and Schr{\o}der T~B 2007
  \emph{arXiv:0706.0813}

\bibitem{valdes_mixing_2009}
Valdes L, Affouard F, Descamps M and Habasaki J 2009 \emph{J. Chem. Phys.}
  \textbf{130} 154505

\bibitem{appignanesi}
Appignanesi G~A, {Rodriguez Fris} J~A, Montani R~A and Kob W 2006 \emph{Phys.
  Rev. Lett.} \textbf{96} 057801

\bibitem{appignanesi_reproducibility_2006}
Appignanesi G~A, {Rodriguez Fris} J~A and Frechero M~A 2006 \emph{Phys. Rev.
  Lett.} \textbf{96} 237803

\end{thebibliography}


\end{document}